\documentclass[fleqn,10pt]{wlscirep}
\usepackage[utf8]{inputenc}
\usepackage[T1]{fontenc}
\usepackage{mathtools,amssymb}
\title{Ultra-fast Kinematic Vortices in Mesoscopic 
Superconductors: The Effect of the Self-Field}

\author[1]{Leonardo Rodrigues Cadorim}
\author[2]{Alexssandre de Oliveira Junior}
\author[1,*]{Edson Sardella}

\affil[1]{Departamento de F\'isica, Faculdade de Ci\^encias, 
Universidade Estadual Paulista (UNESP), Caixa Postal 473, 
17033-360, Bauru-SP, Brazil}
\affil[2]{Instituto de Física Gleb Wataghin, Universidade Estadual de Campinas, P.O. Box 6165, CEP 13083-970, Campinas, Sao Paulo, Brazil}

\affil[*]{edson.sardella@unesp.br}

\begin{abstract}
Within the framework of the generalized time-dependent
Ginzburg-Landau equations,
we studied the influence of the magnetic self-field
induced by the currents inside a superconducting sample
driven by an applied transport current.
The numerical simulations of the resistive state
of the system show that neither
material inhomogeneity nor a normal contact smaller than the
sample width are required to produce an 
inhomogeneous current distribution 
inside the sample, which leads to the emergence of a
kinematic vortex-antivortex pair (vortex street) solution. 
Further, we discuss the behaviors of 
the kinematic vortex velocity, the annihilation rates 
of the supercurrent, and the superconducting order parameters 
alongside the vortex street solution. 
We prove that these two latter points explain 
the characteristics of the resistive state of the system. 
They are the fundamental 
basis to describe the peak of the current-resistance 
characteristic curve and the location where the 
vortex-antivortex pair is formed.   
\end{abstract}
\begin{document}

\flushbottom
\maketitle
%
%
\thispagestyle{empty}


\section{\label{sec:level1}Introduction}
The Ginzburg-Landau theory of superconductivity
states that, in the presence of an applied current,
a superconducting sample can sustain
homogeneous superconductivity
until the current reaches a critical value. 
In specific, this refers to the Ginzburg-Landau pair-breaking
current density, $J^{GL}_c$,
where the sample transitions
to the normal state.
Moreover, phase-slip phenomena
enable superconductivity
not to be destroyed
at currents greater than $J^{GL}_c$.
These coexist with a voltage difference across the sample
in a resistive state.

This mechanism occurs in both thin filaments 
and wide superconducting film samples. Thin filaments
possess dimensions perpendicular
to the current flow that are much smaller than the 
Ginzburg-Landau coherence length, $\xi$.
The phase-slip occurs at the Phase-Slip Centre (PSC),
where the superconducting order parameter, $\psi$,
periodically reaches zero magnitude
with a phase drop of $2\pi$ \cite{ivlev1984}.
Wide films possess only one dimension that is
less than the coherence length, $\xi$.
The resistive state can be realized using
two different processes: a Phase-Slip Line (PSL) and a vortex street.
A PSL is analogous to the PSC
for two dimensions, where the order parameter and the phase
drop occur at a line perpendicular
to the current flow in the sample.
A vortex street, however, is a state
where kinematic vortices move along a line perpendicular
to the applied current of suppressed superconductivity \cite{weber1991}. 
Although the order parameter is very small along this 
vortex street, its phase carries 
two singularities where $\psi$ is consistently zero. They have been 
experimentally observed by Sivakov \textit{et al.} \cite{Sivakov2003}, who 
measured them using the Shapiro steps under microwave radiation 
produced by annihilating the kinematic vortex-antivortex (V-Av) pairs. 
In addition, kinematic vortices posses different characteristics from 
both Abrikosov and Josephson vortices. In particular, their 
velocities, as investigated theoretically and 
experimentally by Jeli\'c \textit{et. al.} \cite{jelic2016} and Embon \textit{et. al.} \cite{embon2017}, 
can be 
greater than Abrikosov vortices and smaller than Josephson vortices. 

A number of numerical works address
resistive states in wide superconducting films.
Andronov \textit{et al.}
\cite{andronov1993} simulated 
homogeneous and inhomogeneous
wide superconducting films and
encountered both PSL and vortex street
solutions.
In addition, Weber and Kramer \cite{weber1991}
investigated a similar configuration
and provided solutions to a larger set of
initial conditions and sample parameters.
Berdiyorov \textit{et al.}
\cite{berdiyorov2009} followed
the changing states of a
superconducting film while increasing
the applied current. 
They studied the \textit{I-V} characteristics
of the sample, as well as the velocity
and nucleation/annihilation position
of the pair of kinematic vortices. They also 
investigated the influence of a
perpendicular applied magnetic field on
these physical quantities.
He \textit{et. al.} \cite{he2016, he2017}
considered the effects of narrow slits
inside the superconducting film
and followed the behavioral changes
of both kinematic vortices and PSLs on such systems. By
varying the size and angle of the narrow slits,
they encountered several different
configurations when increasing
the applied current.
In a different system,
Xue \textit{et al.} \cite{xue2017}
studied the effects of radially injected currents
on a square superconducting film containing a square slit
at its center. 
They found that the current caused the kinematic vortices
to rotate around the square, inducing a voltage oscillation.
The increased external current
motion of the vortices depended on the magnitude
of the applied field.
When investigating a finite superconducting stripe,
Berdiyorov \textit{et. al.}
\cite{berdiyorov2009b} found that an increase in the
$\gamma$ parameter, which is proportional to 
inelastic collision time, caused the phase slip process
to occur in a larger current range. 
In addition, for large values of $\gamma$,
a small applied magnetic field increased the critical current at which the system transits to the normal state. 
Moreover, heat dissipation on the resistive state contributed to quantitative
changes in the size of voltage jumps
and the value of the critical currents.
However, it did not lead to any new qualitative features
\cite{elmurodov2008, vodolazov2005}.
Lastly, Barba-Ortega \textit{et al.}
\cite{barba2019} investigated the influence of a sample's rugosity and found that it influenced both the critical
currents and the kinematic vortex velocity.

To the best of our knowledge,
none of the works in the present
literature have considered
the effects of the magnetic
self-field induced by the internal
currents in the superconducting sample
to study the behavior of ultra-fast kinematic vortices.
The reader should note, though, that this have been done for slowly moving Abrikosov vortices \cite{berdiyorov2012}.
The aim of this paper 
is to show that, although small,
the effects of the self-field are
not negligible and produce
important consequences to the
resistive state, specifically the 
dynamic of the kinematic V-Av
nucleation and annihilation, 
and the peaks present in the resistive characteristic curve.


\section{\label{sec:level2}Results and Discussion}
\begin{figure*}[!ht]
	\begin{center}
		\includegraphics[width=0.9\textwidth]{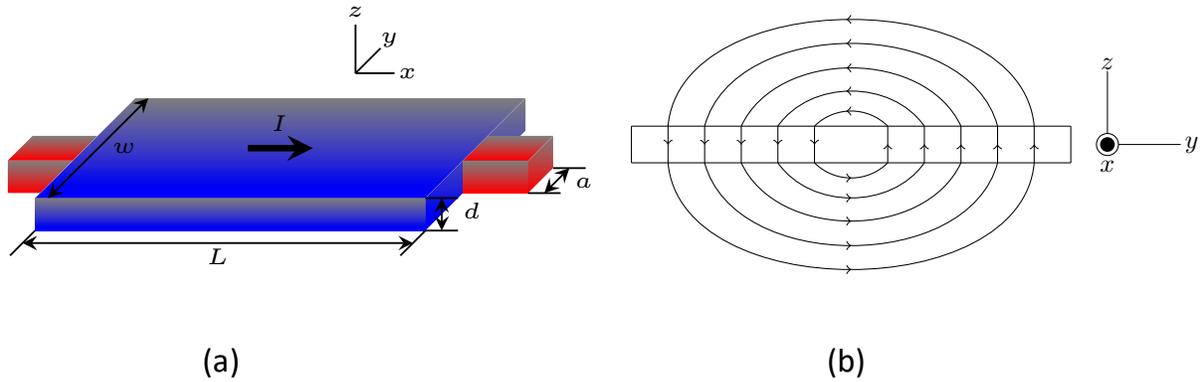}
		\caption{(Color online) (a) A schematic view of a 
		mesoscopic superconducting stripe. The metallic 
		contacts are attached to both sides of the film, 
		through which an external applied transport current, 
		$I$, is injected. 
		The dimensions are indicated 
		in the figure. (b) According to the Amp\`{e}re law, 
	    	the transport current yields a self-field with streamlines as illustrated. 
	    	The current flows in the $x$ direction along 
	    	the film.}\label{fig1}
	\end{center}
\end{figure*}

We investigated a system consisting of a stripe attached to 
two metallic contacts on both sides, through which an applied current 
density, $J_a$, was injected. The length and width of the stripe are denoted by $L$ 
and $w$, respectively. The width of the normal contacts is represented by $a$. The 
thickness is represented by $d\ll \xi,\lambda$, while $\lambda$ represents
the field penetration depth. 
Fig.~\ref{fig1}(b) illustrates the local 
magnetic self-field produced by the applied current. 
The local magnetic field was assumed to be 
perpendicular to the stripe. The 
validity of this approximation is discussed 
in more detail in Section \ref{sec:level3} 
and in the supplementary material.

We considered mesoscopic superconducting stripes
of dimensions $12\xi\times 8\xi$.  
We assumed that the size, $a$,
of the normal contact responsible for the injection
of current in the superconducting stripe was equal to the sample
width, $a = 8\xi$. The Ginzburg-Landau parameter, $\kappa$, and the constant, $\gamma$, were assumed to be $5.0$ and $20$, respectively. Note that, given the thin film geometry, this is an effective $\kappa$ value, which depends on the thickness of the stripe\cite{milovsevic2010}. 
In most cases, the results were displayed as $I=aJ_a$ 
(the total current injected per unit length) rather than $J_a$. 
In the numerical simulations, we adiabatically
increased the transport current in steps of
$\Delta I = 0.079 I_0$
until the whole sample reached the normal state.
In all calculations, the external magnetic
field was $H=0$; here $I_0=\xi J_0$ 
(see Section \ref{sec:level3} for 
the definition of $J_0$, just after equation (\ref{eq:pot})). 
The boundary conditions for the local magnetic field 
did not account for the external applied field since 
we are interested exclusively in the effect of the the self-field (see Methods for details of the theoretical formalism used). 

\begin{figure}[!h]
	\centering
	\includegraphics[width=0.5\textwidth]{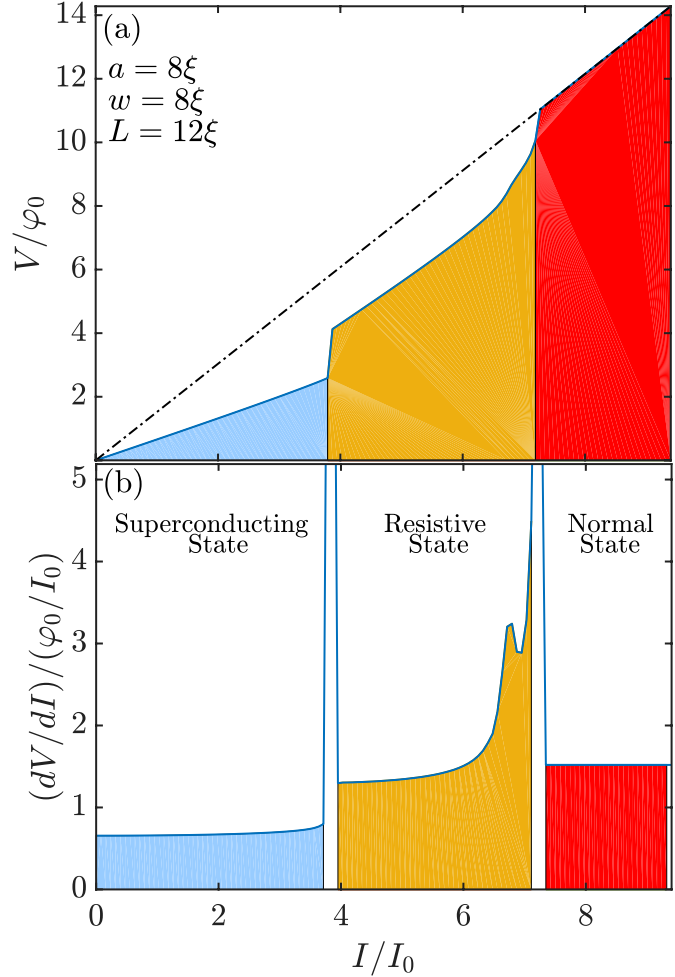}
	\caption{(Color online) (a) The \textit{I-V} 
	characteristics of the system with a normal 
	contact of size $a=8 \xi$. (b) The 
	differential resistance for the same 
	system. The blue, yellow, and red 
	regions represent the Meissner, 
	resistive, and normal states, 
	respectively.}\label{fig:fig3}
\end{figure}

In Fig.~\ref{fig:fig3}(a)
we show the current-voltage characteristics for the system.
For currents where $I < 3.792 I_0$,
the superconducting sample was in 
the Meissner state, without any 
dissipation process: the finite 
voltage presented is caused solely by the normal contacts.
At $I = 3.792 I_0$, a voltage step occurred in the \textit{I-V}
characteristics and the system went into a resistive state.
This resulted in the formation of a vortex street with suppressed superconductivity
at the center of the sample and perpendicular
to the applied current,
where a pair of kinematic V-Avs 
moved from the edges towards the center of the sample.
This is the first manifestation
of the effects of the self-field, since when $a=w$, 
simulations neglecting the self-field
reported in the literature
\cite{berdiyorov2009, weber1991}
showed that the vortex street did not occur. Instead, 
the resistive state found in these cases is 
characterized by a time periodic formation of 
PSLs at the center of the sample. This 
remarkable difference to the results of 
investigations which disregarded the self-field shows 
the importance of such effect to the resistive state.

The currents induced by the self-field
are responsible for enhancing the inhomogeneity
of the supercurrent distribution along 
the width of the sample,
which are otherwise approximately uniform
\cite{berdiyorov2009}. This break of homogeneity
was responsible for the formation of a PSL
for the initial parameters.
It favors a solution that is now dependent on the $y$
coordinate: the vortex street solution.
This does not prohibit the existence of the PSL solution for
smaller samples
where the current distribution is more homogeneous.

As previously mentioned, in the resistive state, kinematic V-Av pairs
were created at the borders and moved towards the center
of the sample. This is another effect of the
self-field,
since simulations without its consideration, as was also
reported in the literature \cite{berdiyorov2009},
presented the same voltage step and transition
to the resistive state with a vortex street solution; however,
the kinematic V-Av pairs 
presented a behavior opposite
to the one described above:
the pairs were nucleated at the center of the sample
and annihilated at its edges. 
This change is also linked to the current density modified by the self-field, more specifically, to the changes it produces in the supervelocity.

\begin{figure}[!h]
	\centering
	\includegraphics[width=0.75\textwidth]{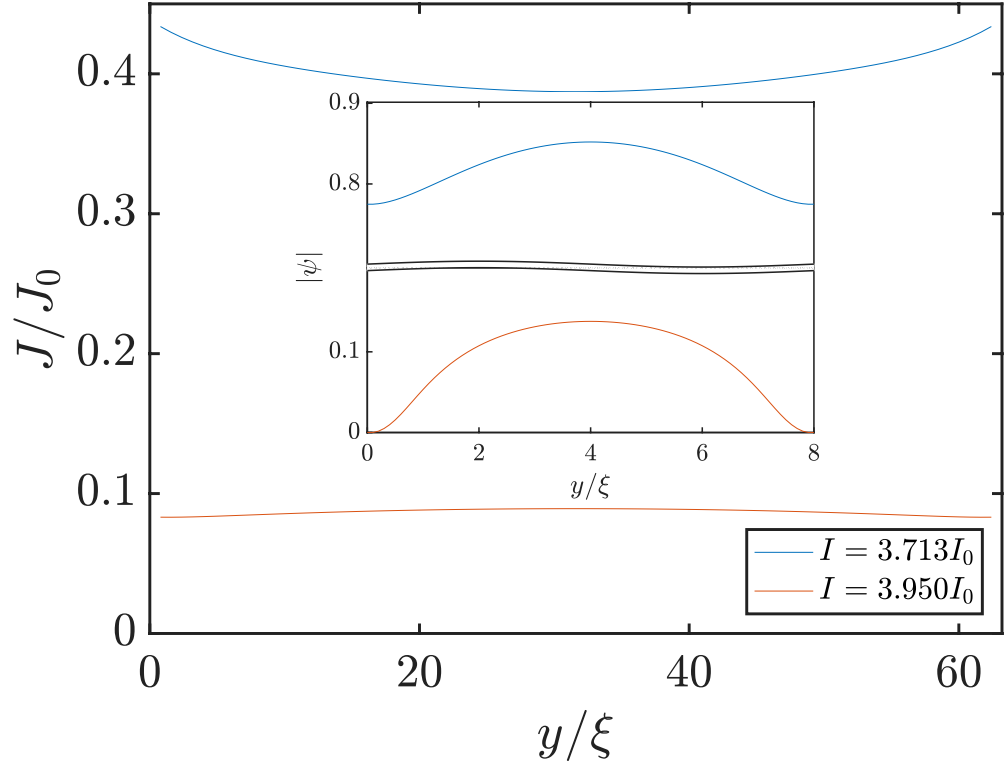}
	\caption{(Color online) The superconducting 
	density current across the width of the 
	sample, at its center, for two different 
	applied current values, $I=3.713 I_0$ 
	(blue curve) and $I=3.950 I_0$ (red curve). 
	The inset shows the superconducting order 
	parameter across the width of the sample.}\label{fig:fig5}
\end{figure}

The supervelocity, which can be expressed as $\textbf{v}=\textbf{J}_s/{\vert\psi\vert}^2$, has its highest value at the point where a vortex nucleates in the sample.
For cases without a self-field,
the supervelocity had its highest value,
for current values right after the first step in voltage,
at the center of the sample \cite{berdiyorov2009}.
On the other hand,
when the self-field was properly considered in the simulations,
the supervelocity had its highest value at the edges,
as shown in Fig.~\ref{fig:fig5}.
Here the supercurrent density distribution is approximately uniform,
but the order parameter is much more suppressed
near the borders of the sample,
resulting in maximum values for supervelocity in those regions.
The currents responsible
for the self-field cause the supervelocity
to reach maximum values at the edges rather than at its center. 
This will have an important consequence
on the subsequent phenomena encountered.

When increasing the applied current,
the system remained at this resistive
vortex street solution without significant change until
the current reached $I = 6.794 I_0$,
where the \textit{I-V} characteristics 
curve's slope changed. 
This caused a peak
in the differential resistance,
as shown in Fig.~\ref{fig:fig3}(b).
The same phenomenon was observed for numerical simulations
that did not consider the self-field effects
\cite{berdiyorov2009}. However, in that work,
this was linked to a change in the positions
of the nucleation and annihilation
of the kinematic V-Av pairs,
which were created at the borders of the sample and
annihilated at its center. 
In the self-field simulations,
the creation and annihilation process always 
took place in the latter form. 
This raises the question of what is really responsible
for the slope change in the \textit{I-V} curve 
and the maximum values at the differential resistance.

\begin{figure}[!h]
	\centering
	\includegraphics[width=0.75\textwidth]{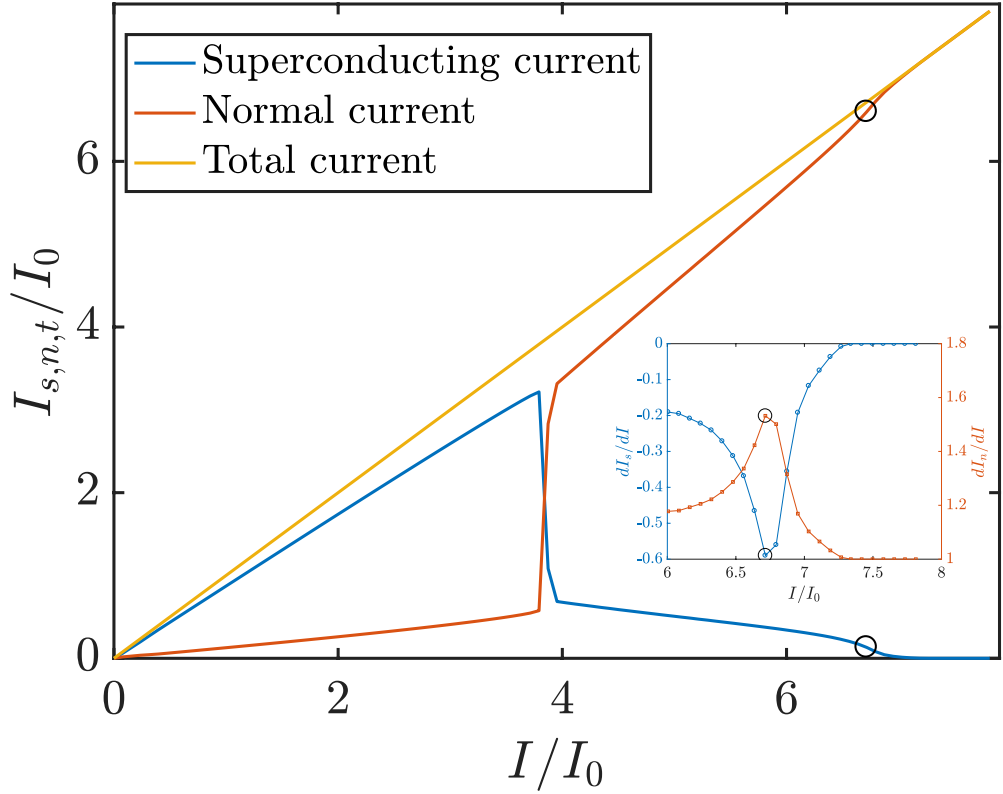}
	\caption{(Color online) Total 
	superconducting current (blue curve) 
	and total normal current (red curve) 
	that cross through the width of the 
	sample, at its center, as functions 
	of the total applied current. The 
	inset shows the rate of change of 
	the superconducting and normal 
	current as functions of of the 
	applied current.}\label{fig:fig7}
\end{figure}

The differential resistance peak found at $I = I_R = 6.794 I_0$ 
was accompanied by other interesting phenomena.
For instance, there was a decrease in the rate at which the superconducting
current was converted to normal current
inside the sample. Fig.\ref{fig:fig7}
shows the total superconducting ($I_s$) 
and total normal ($I_n$) currents that
passed through the whole width of the sample,
at its center, as a function of the total applied current ($I$).
For currents below $I = 3.792 I_0$,
the total superconducting current increased at a
fairly constant rate.
However, at the transition point, $I_s$ dropped abruptly
while $I_n$ increased substantially.
For larger values of the applied current,
the superconducting current
was converted to normal current
at an increasing rate until the applied current reached $I = 6.715 I_0$,
just one step $\Delta I$ smaller than $I_R$, where the differential resistance was maximum. At this point, the rate of conversion 
$dI_n/dI$ reached its maximum value and,
thereafter, decreased with increasing applied current.
Fig.\ref{fig:fig7} shows the rate of change of $I_s$ and $I_n$ as functions of the
applied current $I$. The superconducting current's rate of destruction
reached its maximum at $I = 6.715 I_0$.
For greater values, the superconducting current
was still being destroyed
but at a much lower rate.

\begin{figure}[!h]
	\centering
	\includegraphics[width=0.75\textwidth]{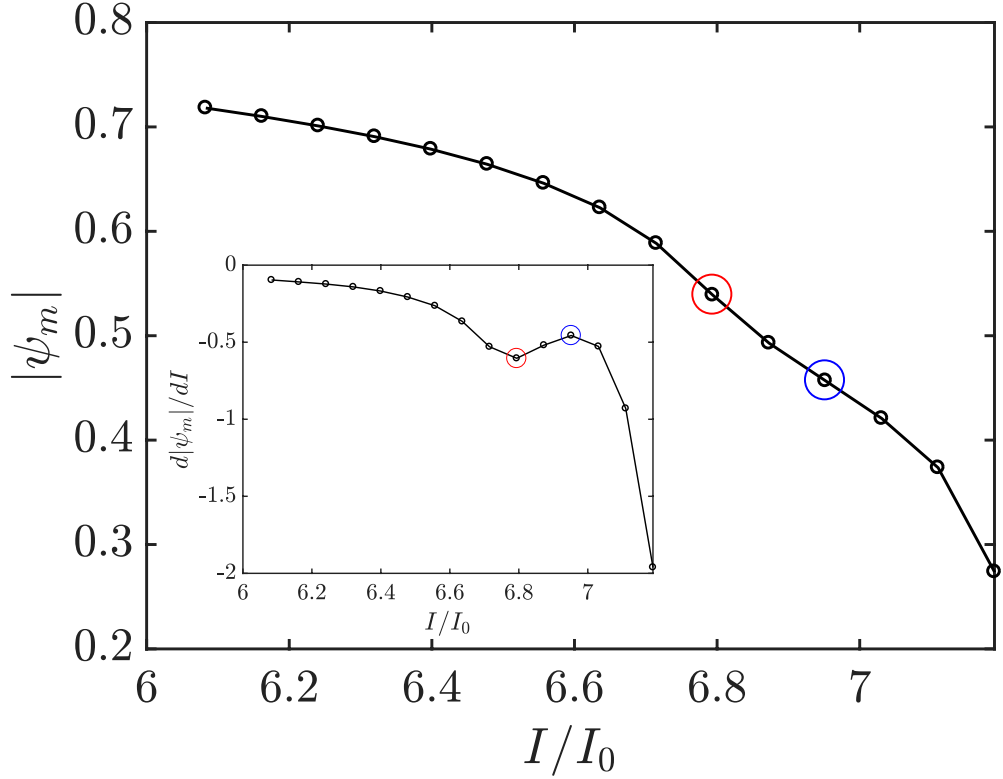}
	\caption{(Color online) The modulus 
	of the time-averaged superconducting order 
	parameter as a function of 
	the applied current. The inset 
	presents the rate of destruction 
	of the order parameter with 
	increasing current. The point 
	marked with a red circle 
	corresponds to $I = 6.794 I_0$ 
	and the one marked with blue 
	circle to $I = 6.952 I_0$.}\label{fig:fig8}
\end{figure}

Another interesting phenomenon that occurred at 
$I=I_R$ was the decreased annihilation rate for the 
superconducting order parameter in the sample.
Fig.~\ref{fig:fig8} shows the modulus of the 
time-averaged superconducting order parameter at the
center of the stripe 
as a function of the applied current. 
The order parameter
monotonically dropped to zero as the current
approached the value of the superconducting-normal
transition,
$I = 7.268 I_0$, as is shown in Fig.~\ref{fig:fig3}. 
The inset of Fig.~\ref{fig:fig8} presents the
rate of change of the time-averaged 
superconducting order parameter 
calculated at the center of the vortex street as 
a function of the applied current.
For currents lower than $I_R$,
the order parameter was annihilated
at an increasing rate. However, for current values greater than $I_R$,
the rate of annihilation decreased until $I = 6.952 I_0$.
This was where the superconducting order parameter
returned to being increasingly destroyed
until the system reached the normal state. These two 
points are highlighted in Fig.~\ref{fig:fig8}.

\begin{figure}[!h]
	\centering
	\includegraphics[width=0.75\textwidth]{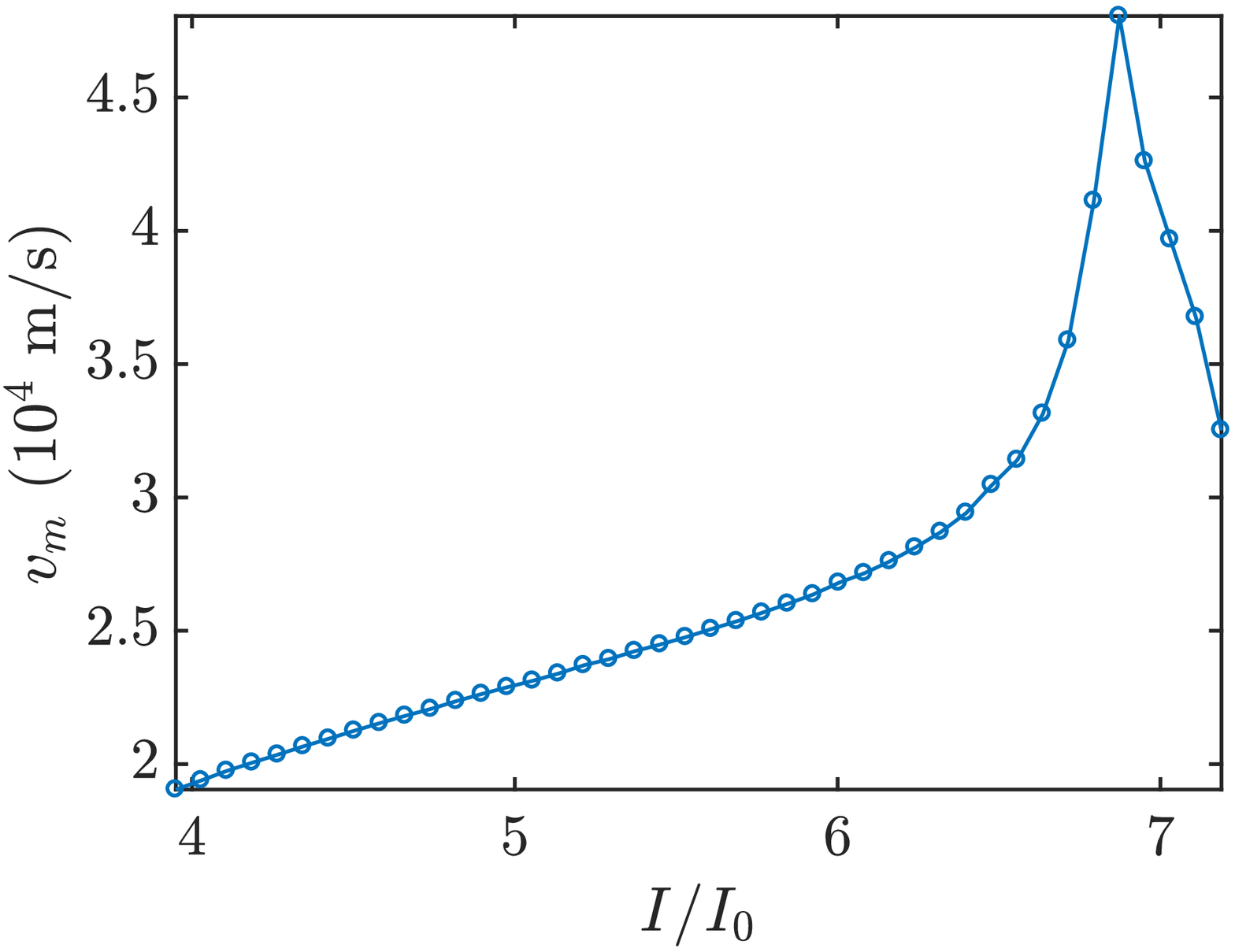}
	\caption{(Colour online) 
	Velocity of the kinematic vortex 
	as a function of the total 
	applied current. The velocity is in real 
	units; we have used $\xi=10$ n$\mu$ 
	and $t_{GL}=6.72$ ps which are 
	typical values for Nb thin films 
	\cite{Gubin2005}.} \label{fig:fig4}
\end{figure}

In these two processes, the decrease 
in the rate of conversion
of the superconducting current to normal current
and the decrease in the rate of annihilation of the
superconducting order parameter in the sample
can be explained by another phenomenon
that took place at $I = 6.873 I_0$, 
just one step $\Delta I$ 
above the $I=I_R$. 
Near this point, the velocity of the kinematic V-Av 
pairs reached its maximum. 
Fig.\ref{fig:fig4} shows the average velocity
of the kinematic vortex as a function 
of applied current.
The average vortex velocity
presented, for currents lower than
$I = 6.873 I_0$, a monotonically increasing
pattern for increasing applied current,
with yet a larger rate of increase for
currents near this value.
However, this tendency to increase
abruptly ceased when the applied current
reaches $I = 6.873 I_0$,
where the kinematic vortex velocity
was maximum.
For higher values,
the average velocity began to decrease
with increasing applied current.

The quasiparticle spectrum changes from superconducting
to normal current when a vortex travels across the sample. 
Thus, with a higher vortex velocity,
the quasiparticles switch more rapidly,
causing an increase in the rate that superconducting current 
is being conversed to normal current and an increase in 
the annihilation rate of the superconducting order parameter.
On the order hand, for applied currents
larger than $I = 6.873 I_0$,
the vortex velocity
starts to decrease, consequently causing a reduction
in both the conversion rate and annihilation rate.
 
Furthermore, we have also encountered that the self-field influences the magnitude of the 
kinematic vortex velocity in the numerical simulations. 
The average velocity of the kinematic vortex
inside the sample remains finite for all values
of the applied current, as seen in Fig.~\ref{fig:fig4}.  
This is unlike the average velocity obtained in simulations 
with the absence of the self-field \cite{berdiyorov2009}, 
which diverge to infinity at $I=I_R$.

To summarize, in this manuscript, we have numerically 
solved the generalized time-dependent Ginzbug-Landau equation
equation and investigated the resistive state
for a superconducting stripe driven by an applied
transport current.
Contrary to previous literature,
the calculations have explicitly considered the magnetic self-field
induced by the internal currents.
We found that the self-field influences
the density of the superconducting current
by changing the location of the creation and annihilation
of the kinematic V-Av pair.
The self-field can also alter the type of resistive state
found for a given set of geometrical parameters. For instance, 
a system with a normal contact equivalent to the stripe width, 
with $\kappa=5.0$ and $\gamma=20$, 
changes from a PSL resistive state
to a vortex street solution when the self-field
is included in the simulations. 
In addition, we also investigated the influence of the
kinematic vortex velocity in the resistive state.
The results maintained that, above certain applied current values, 
the vortex velocity ceases to increase and begins
to decrease, subsequently decreasing the rate at which
that the superconducting current is converted to normal
current and decreasing the annihilation rate
of the superconducting order parameter.
Finally, our results show that the self-field
has important consequences to 
the dynamics of the resistive
state. Thus, it cannot 
be disregarded in similar numerical 
simulations.

\section{\label{sec:level3}Methods}
We have used 
the generalized time-dependent Ginzburg-Landau 
equation (see references \cite{kramer1978} 
and \cite{tobin1981}). In dimensionless form, 
this equation can be written as
\begin{eqnarray}\label{eq:gtdgl}
    \frac{u}{\sqrt{1+\gamma^2|\psi|^2}}
    \left (\frac{\partial}{\partial t}
    +\frac{\gamma^2}{2}\frac{\partial|\psi|^2}
    {\partial t}+i\varphi\right )\psi & \nonumber \\
    = -\left
    (-i\mbox{\boldmath $\nabla$}-{\bf A} \right
    )^2\psi+\psi(1-|\psi|^2)\;. &
\end{eqnarray}
The vector potential was determined using the 
Amp\`ere-Maxwell equation
\begin{equation}
	\frac{\partial{\bf A}}{\partial
	t}+\mbox{\boldmath $\nabla$}\varphi =  {\bf
	J}_s-\kappa^2\mbox{\boldmath $\nabla$}\times
	{\bf h}\;.\label{eq:AME}
\end{equation}
where the superconducting current density is
\begin{equation}\label{eq:cd}
    {\bf J}_s = {\rm Re}\left [ \bar{\psi}\left (
    -i\mbox{\boldmath $\nabla$}-{\bf A} \right )\psi
    \right ]\;,
\end{equation}
and the local magnetic field is related to the vector potential 
through the equation 
$\textbf{h}=\mbox{\boldmath $\nabla$}\times\textbf{A}$.

The equation for scalar potential can be derived 
from the continuity equation
\begin{equation}
    \frac{\partial \rho}{\partial t}
    +\mbox{\boldmath $\nabla$}\cdot\textbf{J}=0\;,
\end{equation}
where $\textbf{J}=\textbf{J}_s+\textbf{J}_n$, and
\begin{equation}
    \textbf{J}_n=-\left(\frac{\partial{\bf A}}
    {\partial t}+\mbox{\boldmath $\nabla$}\varphi\right)\label{eq_gtdgl_dcn}
\end{equation}
is the normal current density. Supposing that there 
is no accumulation of charge, we can write 
$\frac{\partial \rho}{\partial t}=0$, which 
yields $\mbox{\boldmath $\nabla$}\cdot\textbf{J}=0$. 
Then, by assuming the Coulomb gauge 
$\mbox{\boldmath $\nabla$}\cdot{\bf A}=0$, from 
(\ref{eq:AME}) we can easily obtain
\begin{equation}
    \nabla^2\varphi = \mbox{\boldmath $\nabla$}
    \cdot{\bf J}_s\;.\label{eq:pot}
\end{equation}

Here, lengths are in units of the coherence 
length, $\xi$,  temperature, $T$, is in units of $T_c$, 
and time is in units of the GL 
time characteristic 
$\tau_{GL}=\pi\hbar/8k_BT_c\epsilon u$, where $\epsilon=(T_c-T)/T_c$. In addition, the magnetic field 
is in units of the upper critical field, $H_{c2}$, 
the electrostatic potential is in units of 
$\varphi_0=\hbar/2e\tau_{GL}$, the vector potential is 
in units of $H_{c2}\xi$, the current density is in units 
of $J_0=c\sigma\hbar/2e\xi\tau_{GL}$ 
(where $\sigma$ is the electrical 
conductivity in the normal state), and the 
order parameter is in units of $\psi_0 = \sqrt{|\alpha|/\beta}$ 
(the order parameter in the Meissner state). Lastly, 
$\alpha$ and $\beta$ are the GL phenomenological 
constants, and $\kappa=\lambda/\xi$ 
is the Ginzburg-Landau parameter, and  
$\lambda$ is London penetration length. 
The constant $u=5.79$ was derived from the first 
principles in Refs.~\citeonline{kramer1978,tobin1981}.

We have solved equations (\ref{eq:gtdgl}), (\ref{eq:AME}), and 
(\ref{eq:pot}) numerically for the geometry exhibited in 
Fig.~\ref{fig1}(a). Along all sides 
of the film, $\textbf{n}\cdot\mbox{\boldmath $\nabla$}\varphi=0$ 
except on the normal contacts where 
$\textbf{n}\cdot\mbox{\boldmath $\nabla$}\varphi=-J_a$. In the limit 
of thickness $d\ll\xi$ inside the film, we may 
consider that the self-field was nearly 
perpendicular to the film along the $z$ direction. The validity 
of this approximation has been rigorously proved 
in Ref.~\citeonline{Chapman1996} to be good for large $\kappa$, typically 
$\kappa \gtrsim 5$ \cite{Du1995}. Thus, within 
this approximation, the boundary conditions for the self-field are 
\begin{eqnarray}
h_z\bigg(x,\pm\frac{w}{2}\bigg) &=& \pm\frac{J_aa}{2\kappa^2}\;, \nonumber \\
h_z\bigg(\pm\frac{L}{2},y\bigg) &=& 
\begin{dcases}
\dfrac{J_aa}{2\kappa^2}\;,     &  \dfrac{a}{2} \le y \le \dfrac{w}{2}\;, \\
\dfrac{J_ay}{\kappa^2}\;, & -\dfrac{a}{2} \le y \le \dfrac{a}{2}\;, \\
-\dfrac{J_aa}{2\kappa^2}\;,     &  -\dfrac{w}{2} \le y \le -\dfrac{a}{2}\;,
\end{dcases}
\end{eqnarray}
which can be easily obtained from the Amp\`ere's 
law  $\oint\,\textbf{h}\cdot d\textbf{l}=Id/\kappa^2$ 
in dimensionless units; here $I=\int J_x(y)\, dy$. Since the thickness 
of the film is very small and homogeneous, then $h_z$ does 
not depend on $d$. This 
same assumption has been used in, for instance, Ref.~\citeonline{berdiyorov2012}. 
The order parameter is determined 
at the border of the sample 
using the Neumman boundary condition 
$\textbf{n}\cdot\left (-i\mbox{\boldmath $\nabla$}
-{\bf A} \right )\psi=0$ which assures that the 
perpendicular component of the superconducting 
current density vanishes at all sides of the sample.

We solved the equations upon using the 
link-variable method as sketched in reference \citeonline{gropp1996}. 
The equations were discretized in a mesh-grid of 
size $\Delta x=\Delta y=0.1\xi.$

As a final remark on our method, we emphasize that our approximation depends on the smallness of the $h_y$ component of the magnetic field in our sample. In the supplementary material, we argue that, for the geometry under investigation, this indeed occurs.


\bibliography{sample}


\section*{Acknowledgements}

LRC thanks the UNESP for financial support. ES thanks 
the Brazilan Agency FAPESP for financial support 
(process number 12/04388-0). ADJ thanks 
FAPESP for grants (process number 15/21189-0). 

\section*{Author contributions statement}

LRC and ADJ carried out the numerical simulations. All authors helped write the manuscript. ES supervised and conceived the problem. 

\section*{Additional information}



The authors declare no competing interests.

\section*{Availability}
 The code used to carry out the numerical simulations 
 can be obtained freely by request to the corresponding author.

\end{document}